# Kinetic studies on using photocatalytic coatings for removal of indoor volatile organic compounds

Zhuoying Jiang[*], Xiong (Bill) Yu[*]

Department of Civil Engineering, Case Western Reserve University, Cleveland, OH 44106, United States

[*]Corresponding authors: Email: xxy21@case.edu (X. Yu), zxj45@case.edu (Z. Jiang)

**Abstract**

Titanium dioxide ($TiO_2$) is a known photocatalyst with a capability of decomposing organic substance. However, the photocatalysis of the pure $TiO_2$ is not effective for the indoor environment due to a lack of the ultraviolet irradiation inside the building. Doping $TiO_2$ with substance such as C, N or metal can extend the threshold of the absorption spectrum to the visible spectrum region. Thus, doped-$TiO_2$ is able to decompose volatile organic compounds (VOCs) under an indoor environment. To date, most experimental work reported on photocatalytic kinetics were conducted inside small-scale devices. The performance of air purification function under the actual indoor application scenery need to be further clarified. For this purpose, it is crucial to predict the performance of autogenous air quality improvements by visible light driven photocatalyst for the actual applications. This work has developed a model to evaluate the performance of functional coating with photocatalyst in removing VOCs. Factors such as the effects of coating designs and indoor ambient conditions on the air purification efficiency were studied. This work demonstrates that doped-$TiO_2$ photocatalytic coating is effective to improve the indoor air quality.

**Keywords:** Volatile organic compounds; doped titanium dioxide; photocatalysis; indoor air quality control.

**Summary of the practical implications**

The kinetics of doped-$TiO_2$ on removal of VOCs are simulated in the actual size room, rather than that in the small-size devices. This work demonstrates that doped-$TiO_2$ photocatalytic coating is effective to remove the volatile organic compounds with a reasonable dosage of photocatalyst at a comparable rate to ventilation. We believe this work fills a knowledge gap for removing the indoor air contaminants from the large-scale dimensional aspect. It should be feasible to fabricate commercial membranes coated with doped-$TiO_2$ for indoor air purification.

## 1. Introduction

With rapid development of modern civilization, air pollution has become an urgent issue, which strongly affects the physical and psychological health of residents. Thus, the improvement of the indoor air quality has received considerable attention [1-3]. The typical indoor air contaminants include particulate matters (PM), volatile organic compounds (VOCs), and bio-aerosols [4,5]. The major contaminants inside the building are VOCs, which are referred to the carbon-contained organic substances in the air. VOCs are usually not acutely toxic, but it causes an adverse health effect when human are exposed to a high concentration of the VOCs for a long time. Formaldehyde, toluene, and dichloroethane, etc., are common VOCs in the indoor environment [6,7]. Recent studies have found that VOCs are considered to be one of the major causes of the sick building syndrome (SBS) [8], which identifies a condition that the building occupants experience discomfort or even illness during their stay inside the building [9]. The discomfort includes headache, eye, nose and throat irritations, cough, itchy skin, dizziness and nausea. The traditional approach to remove the air contaminants and maintain a clean indoor environment is by the ventilation. However, SBS usually occurs in the working environment with a poor ventilation.

Therefore, the photocatalytic coating is proposed in this work to be an alternative approach to facilitate the VOC removal, especially for a working environment with a poor ventilation.

A number of VOCs have been reported to be effectively decomposed by $TiO_2$ [10-12]. Besides, this photo-decomposition is non-selective, i.e., VOCs with different chemical structures and functional groups, including aliphatics, aromatics, alcohols, ethers, ketones, halogen hydrocarbons, can mostly be decomposed by $TiO_2$ [10,11]. However, the absorption spectrum of the $TiO_2$ is limited in an ultraviolet band due to its large bandgap (3.2 eV). Since there is little ultraviolet content in an indoor lighting environment, pure $TiO_2$ can be hardly activated inside the building. Doped-$TiO_2$ materials exhibits a capability to be activated by visible light [13]. The artificial $TiO_2$ with impurities can extend its threshold of the absorption spectrum to the visible region with a narrower band-gap. Consequently, the doped-$TiO_2$ is able to maintain its photocatalytic functionality and decompose the VOCs in an indoor lighting environment. Typically, there are two types of doped impurity elements in $TiO_2$: metals and non-metals. Metals includes manganese, iron, cobalt, and nickel [14,15], and non-metals include boron, carbon and nitrogen [16-19]. The metal and non-metal elements can also be co-doped in $TiO_2$, e.g., iron-nitrogen co-doped [20], which helps to further reduce the band-gap and improve the photocatalytic activity.

Since it is capable of decomposing VOCs, the doped-$TiO_2$ photocatalytic coating is considered as a promising candidate for the indoor air purification. The room lights, which are usually in the visible light range, can be used as the activation energy for the doped-$TiO_2$ coating with no additional cost. Most experimental work on photocatalytic efficiencies reported was in a small-scale device [21-25]. Prior to applications, assessment of the removal efficiency for indoor environment, which is typically much larger in volume than the small-scale devices used in proof-of-concept study, is important to determine the feasibility. This can be achieved with the kinetic studies to quantify the air cleaning efficiency of the doped-TiO$_2$ coatings. For this purpose, this work describes the development and application of a computational model to analyze the efficiencies of nitrogen doped-TiO$_2$ (N-TiO$_2$) photocatalytic coatings in decomposing VOCs.

## 2. Modeling approach

### 2.1 Description of the 3D model

A three-dimensional model was built to investigate the kinetics of indoor VOCs removal with the presence of the N-doped TiO$_2$ photocatalytic coatings. The geometry and dimensions are shown in Fig. 1 (a) and Table 1, respectively. Areas where the photocatalytic coatings are applied are indicated in blue color, as shown in Fig. 1 (b) and (c). The total coated areas in Fig. 1 (b) and 1 (c) are kept the same as 6.6 m$^2$. The effect of coating layouts on VOC removal efficiency will be investigated. In addition, to compare the efficiency of removing VOCs between using photocatalytic coatings and using ventilation system, this model also incorporates the ventilation function, and was built as shown in Fig. 1 (d). The length and width of the ventilation window are assumed to be 0.5 m and 0.25 m respectively. The speed of inlet clean airflow is set to be 0.01 m s$^{-1}$ according to the American society of heating, refrigerating and air-conditioning engineers (ASHRAE) standards [26].

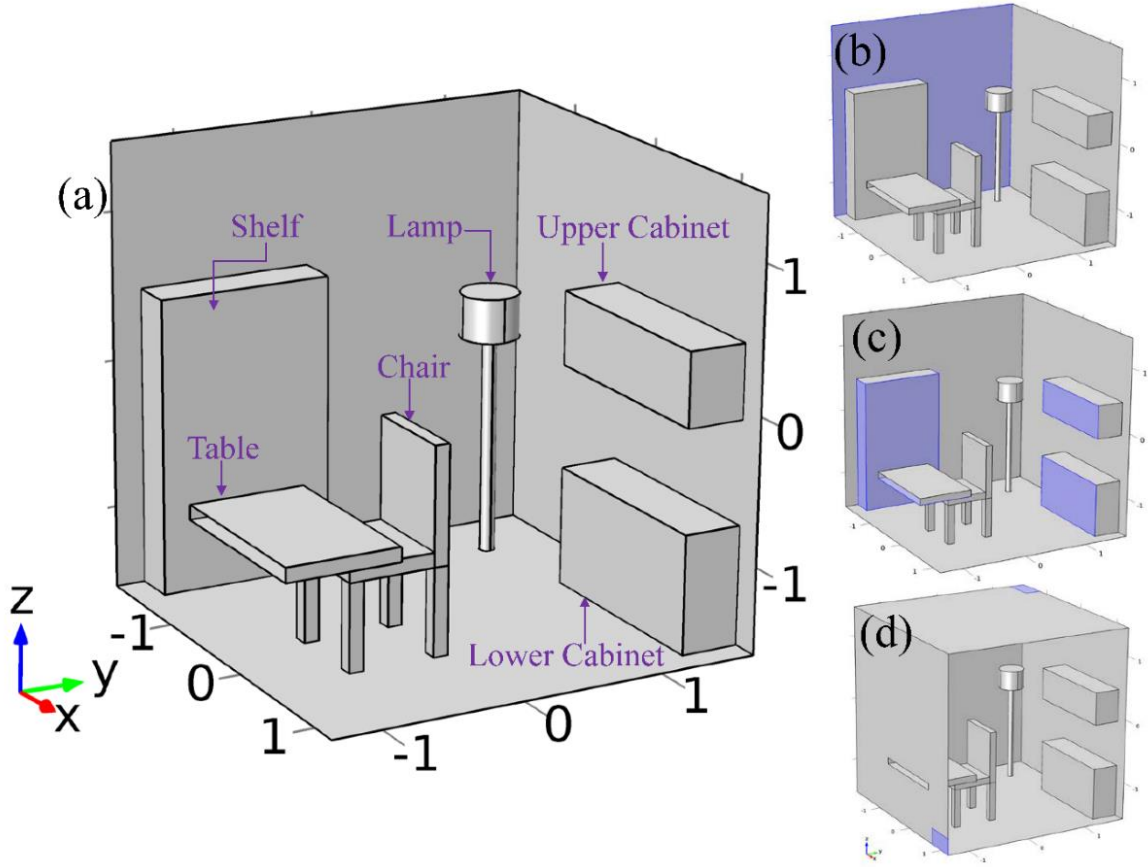

Figure 1: (a) Three-dimensional geometry of the model, (b) areas of walls covered with photocatalytic coating (indicated with blue color) on one wall surface, (c) photocatalytic coatings applied on several separate wall surfaces with a total area equals to that in (b), and (d) ventilation system.

Table 1: Dimensions and positions of the room and each furniture.

| Components | | Size (m) | | | Coordinates of the central position |
|---|---|---|---|---|---|
| | | Width (x) | Depth (y) | Height (z) | (x, y, z) (m) |
| Room | | 3 | 3 | 3 | (0, 0, 0) |
| Shelf | | 0.3 | 1.2 | 2 | (-1.35, -0.7, -0.5) |
| Table | | 1.3 | 0.8 | 0.1 | (0.25, -1.1, -0.65) |
| Chair | Back | 0.7 | 0.1 | 0.8 | (0.25, -0.25, -0.5) |

| | Seat | 0.7 | 0.6 | 0.1 | (0.25, -0.6, -0.85) |
|---|---|---|---|---|---|
| | 4 Legs | 0.1 | 0.1 | 0.6 | (-0.05, -0.85, -1.2), (-0.05, -0.25, -1.2), (0.55, -0.85, -1.2), (0.55, -0.25, -1.2) |
| Upper cabinet | | 1.5 | 0.4 | 0.5 | (0.55, 1.3, 0.25) |
| Lower cabinet | | 1.5 | 0.4 | 0.8 | (0.55, 1.3, -1.1) |
| Lamp | Lamp shade | Radius: 0.2 | | 0.3 | (-1, 1, 0.15) |
| | Lamp base | Radius: 0.05 | | 1.5 | (-1, 1, -0.75) |

To accelerate the removal of the VOCs, a fan was built in the room, as shown in Fig. 2. The size and position of the fan are provided in Table 2. With a fan blowing, forced convection accelerates the migration of VOCs inside the room. To simplify the model with a ventilation system, the furniture are not considered in the model. The area of the coating surface is the same for both the model with the fan (without the furniture) and the model without the fan (with furniture).

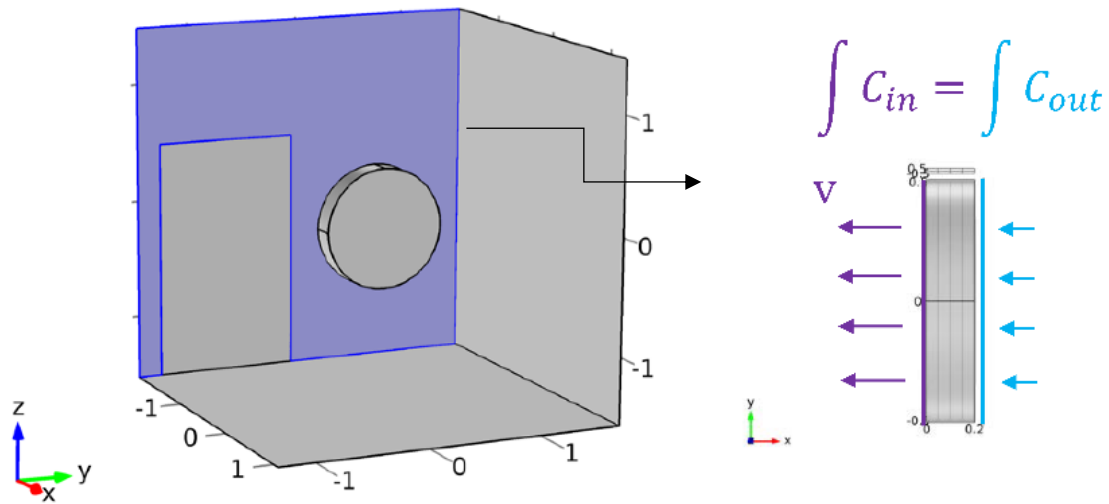

Figure 2: The geometry of the model with the fan, unit: m.

Table 2: Dimensions and positions of the room and the fan.

| Component | Dimension (m) | | | Coordinates of the central position (x, y, z) (m) |
|---|---|---|---|---|
| | Width (x) | Depth (y) | Heights (z) | |
| Room | 3 | 3 | 3 | (0, 0, 0) |
| Fan | 0.2 | Radius: 0.5 | | (0, 0, 0.1) |

For a given VOC completely decomposed by photocatalysts, the reaction is written as

$$VOC \xrightarrow{Photocatalyst;\ light} Intermediate \xrightarrow{Photocatalyst;\ light} CO_2 + H_2O \qquad (1)$$

The VOCs removal efficiency by the photocatalytic coating is controlled by two factors: (1) photodegradation rate of photocatalyst decomposing VOCs, and (2) migration rate of VOCs onto the surface of the photocatalytic coatings. The bulk photodegradation rate is determined as [25]

$$r_{bulk} = kC_{VOC} \qquad (2)$$

Where, $k$ (s$^{-1}$) is the reaction rate constant, and $C_{VOC}$ (mol m$^{-3}$) is the concentration of the VOC on the photocatalyst surface. The reaction rate constant is an inherent property of the photocatalyst for decomposing VOC, and it can be determined by the experiments. In the previous work, the bulk reaction rate was typically used to describe the VOC removal efficiency by a photocatalytic coating in a small-scale device used as the container of the VOC contaminant, with volume ranging 200-2000 mL [22-25]. The effects of diffusion of the VOC under this small size scale can be effectively neglected. Therefore, the reaction rate obtained in the gas container is a bulk equivalent value. However, when the photocatalytic coating is applied in the room, which has a much larger volume than those lab-scale devices, the rate of air movement towards the photocatalytic coating cannot be neglected. The surface reaction rate at the photocatalytic surface is obtained by converting the bulk reaction rate determined from the small-scale experiments.

$$r_{surf} = \frac{V}{A} r_{bulk} = k C_{VOC} \cdot \frac{V}{A} \tag{3}$$

Where, $r$ (mol m$^{-2}$ s$^{-1}$) is the surface reaction rate of the photocatalyst for degrading VOC, $V$ (m$^3$) is the volume of the VOC container, and $A$ (m$^2$) is the coating area of the photocatalyst. In this work, N-doped TiO$_2$ is assumed to be the photocatalyst, and toluene is assumed as the VOC contaminant. Table 3 summarizes the reaction rate constant ($k$) and other parameters, i.e. dosage, light intensity, etc. adopted from reference [25] used in our model. Light intensity has been found to significantly affect the photodegradation rate [27,28]. A higher light intensity means more photons incident onto the photocatalysts, and consequently generates more electrons and holes, which lead to a higher photodegradation rate. In this work, light intensity is assumed to be 0.7 W m$^{-2}$ (equivalent to 500 lux at the average wavelength of 550 nm), which is within the range in a typical study room [29].

Table 3. Parameters used in the model, adopted from literature [25].

| VOC | Photocatalyst | Dosage (g m$^{-2}$) | $k$ (s$^{-1}$) | Light source & intensity (W m$^{-2}$) | $A$ (cm$^2$) | $V$ (cm$^3$) |
|---|---|---|---|---|---|---|
| Toluene | N-doped TiO$_2$ | 2.1 | 1.008×10$^{-5}$ | LED, 0.7 | 1178 | 1000 |

The governing equation for determining the remaining concentration of VOCs based on the mass conservation yields

$$\frac{\partial C_{VOC}}{\partial t} + \nabla \cdot (-D\nabla C_{VOC}) + u \cdot \nabla C_{VOC} = r \quad (4)$$

Where, $D$ (m$^2$ s$^{-1}$) is the diffusion coefficient, $u$ (m s$^{-1}$) is the flow velocity by the forced convection, and $r$ (mol m$^{-3}$ s$^{-1}$) is the photodegradation rate. $r$ is only effective on the photocatalytic surface and is symbolled as $r_{surf}$ as discussed before. In the bulk air, $r$ equals zero. To accelerate the VOC migration onto the photocatalytic coatings, two strategies are considered: (1) accelerate the diffusion velocity, and (2) introduce forced convection towards photocatalytic coatings. In the tranquil air, VOC on the photocatalyst surface begins to decompose with lights turned on. As toluene is consumed on the photocatalytic surface, the toluene in the surrounding starts to diffuse onto the surfaces of the coatings due to the concentration gradients. The effects of diffusion velocity on efficiency of removing VOC under different temperatures are studied. The diffusion velocity is also affected by the room temperature. Table 4 summarizes the diffusion coefficients at different temperatures. This study also examines the effects of the second strategy by bringing in a forced convection as shown in Fig. 2.

Table 4. Diffusion coefficients of toluene at 1 atm.

| Temperature (°C) | Diffusion coefficient ($D$) in air (cm$^2$ s$^{-1}$) | Reference |
|---|---|---|
| 5 | 6.45×10$^{-6}$ | |
| 15 | 7.26×10$^{-6}$ | [30] |
| 25 | 8.04×10$^{-6}$ | |
| 35 | 8.88×10$^{-6}$ | |

Finally, in order to study the photocatalytic functionality for a long term, the releasing rate of VOC from the furniture was also considered in this model. The VOC removal efficiencies were calculated under two releasing rates: a low releasing rate from a carpet, and a high releasing rate from a printer. The surfaces releasing VOC are marked with orange color as shown in Fig. 3, and the releasing rates were assigned referring to the experimental data from Ref [30] and are listed in Table 5.

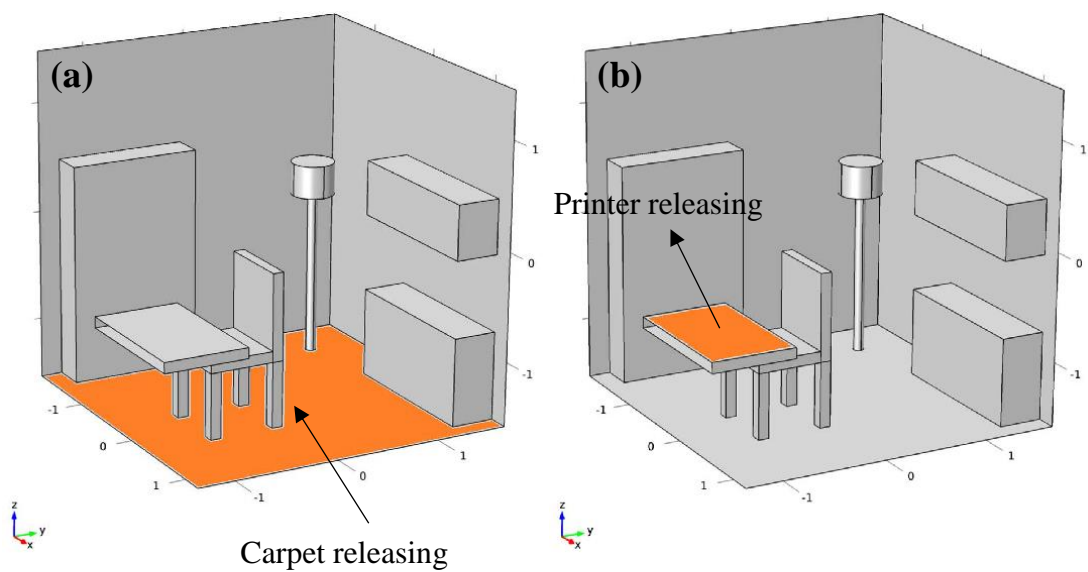

Figure 3: The simulated room with furniture considering VOC releasing. (a) Releasing from a carpet, and (b) releasing from a printer. Releasing surfaces are marked with orange color.

Table 5. Releasing rates used in this work.

| VOC | Releasing sources | Releasing rate ($\mu g\ m^{-2}\ hr^{-1}$) | Reference |
|---|---|---|---|
| Toluene | Carpet | 1.47 | [31] |
| | Printer | 549 | |

The VOC removal efficiency was simulated using COMSOL Multiphysics software package, which has been demonstrated as an effective tool for modeling transport phenomena, i.e. flow mechanics, chemical reactions, electrochemistry, etc [32-36]. This model coupled reaction engineering module, transport of diluted species module and laminar flow module. The relative tolerance of error in the simulation is set as $10^{-2}$ in this model. The mesh independent analysis was also made. For the diffusion model shown in Fig.1, the model with the total number of elements of 10514, 24126 and 51294 yield the same results. Based on this, mesh with 24126 elements are used for the subsequent simulations based on balanced consideration of accuracy and computational efficiency. For the model with circulation fan as shown in Fig. 2, sensitivity study showed that the use of total elements of 329663, 508043, 646398 meshes yield the same results. Therefore, mesh with 508043 total elements are used in the simulation model by considering the balance of accuracy and efficiency.

## 3. Results and discussion

### 3.1 Effect of nitrogen-doped $TiO_2$ dosage

The initial concentration of the toluene in the simulation model is assumed to be 1 ppmv. It was suggested in the previous study that a total level of the VOCs higher than 3.0 mg m$^{-3}$ (~ 0.75 ppmv if the average molecular weight of the VOCs is assumed to be 100 g mol$^{-1}$) in the indoor environment might cause uncomfort or even illness for building occupants [37]. The simulation results of the toluene removal efficiency with one wall and separate surface coatings are shown in Fig. 4. For N-$TiO_2$ dosage of 21 g m$^{-2}$ on the coating, it takes 30 days to remove 40 % of the toluene. The similar removal efficiency is achieved both when coating is applied on one wall and on separate surface coatings with the same total surface area. When N-$TiO_2$ dosage is as high as 1 kg m$^{-2}$, it takes 5 days to remove 80 % of the toluene when photocatalyst is applied to the separate surface coating, while only 60 % of the toluene is removed when photocatalyst is applied to the single wall coating. It was found that for the low N-$TiO_2$ dosage, the removal efficiency is almost the same for two different coating layouts. The toluene concentration is relatively uniformly distributed for each time step. This indicates that diffusion of the toluene is faster than the reaction rate of the toluene decomposition, and therefore the VOC removal efficiency is mainly limited by the reaction rate. Therefore, the removal efficiencies are similar for the same surface areas covered with photocatalysts regardless of how they are distributed. However, for a high N-TiO$_2$ dosage, the toluene concentration is non-uniformly distributed in the room, i.e., lower at the photocatalytic surfaces and higher away from the photocatalytic wall. This means the reaction rate is faster than the diffusion rate, and thus the toluene removal efficiency is limited by the diffusion rate. For the separate surface coatings, the effective diffusion distances are shorter; therefore, the VOC removal efficiency is higher than the case where the coating is applied to the single wall.

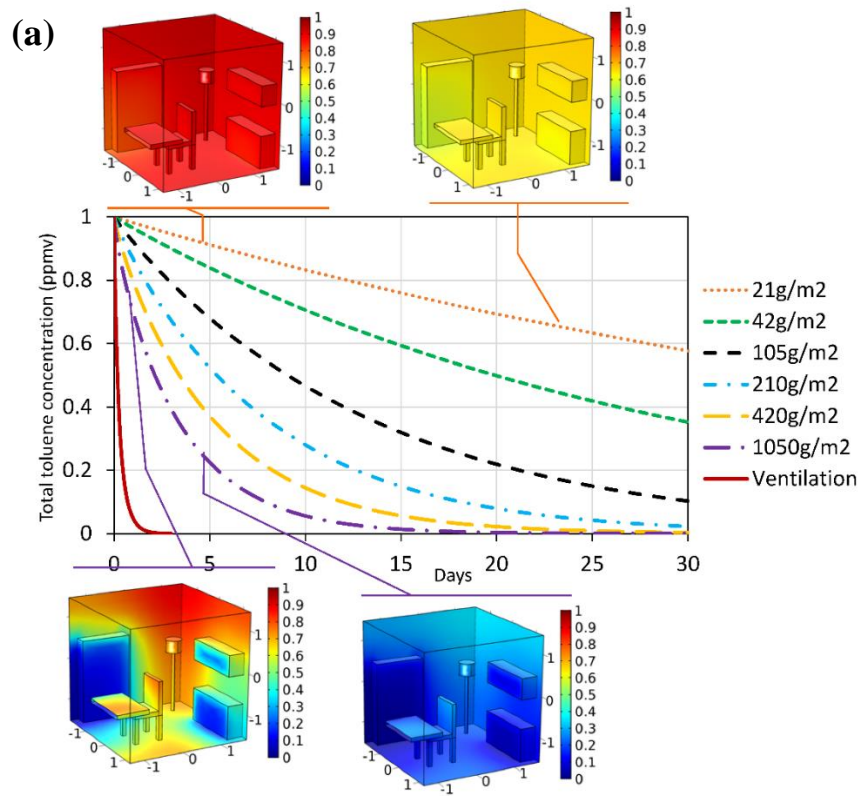

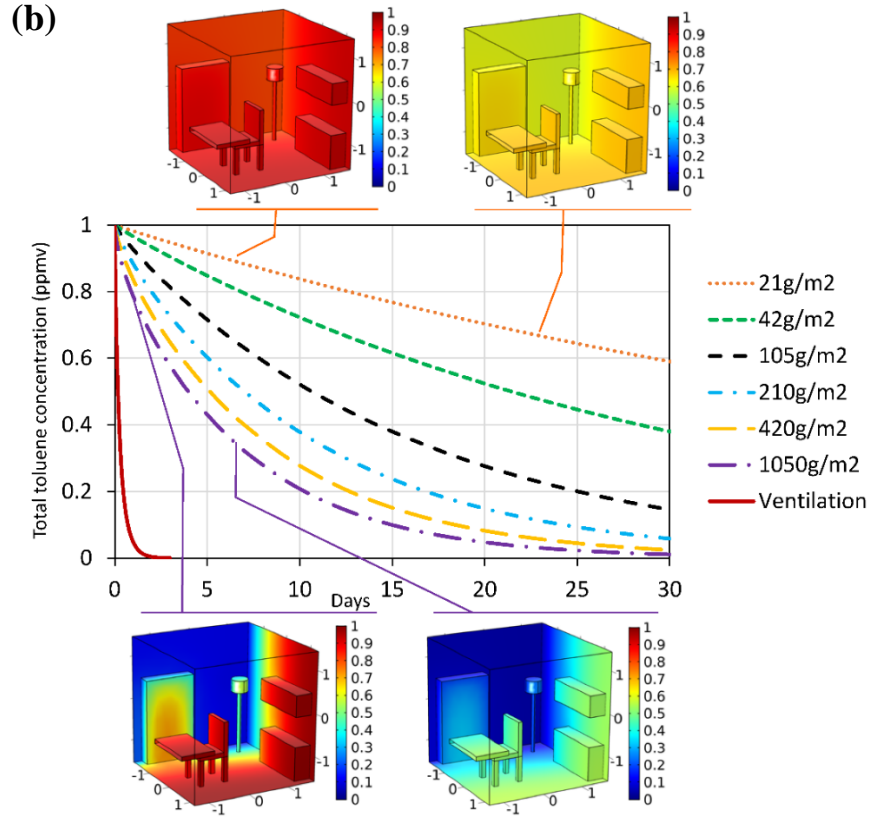

Figure 4: The effects of the N-TiO$_2$ dosage on toluene removal efficiency. (a) Separate surface coatings, and (b) single wall coating.

Moreover, the effects of ventilation in removing toluene are simulated and the results are shown in the red line in Fig. 4. With ventilation, 90 % of the toluene is removed within one day. Compared to the ventilation, the removal efficiency of using photocatalytic coatings is much lower. To enhance the VOC removal efficiency, two strategies are further analyzed: (1) Increase diffusion speed of VOC with a higher room temperature, and (2) add a forced convection with a fan.

**3.2 Accelerating contaminant diffusion by increasing the room temperature**

This section studies the strategy to improve the VOC removal efficiency of N-TiO$_2$ photocatalytic coatings by raising the room temperature to increase the diffusion velocity of VOCs. The effects of room temperature are analyzed, ranging from 5 °C to 35 °C, on toluene removal efficiency are shown in Fig. 5. At a low N-TiO$_2$ dosage, higher temperature does not improve the removal efficiency for both separate surface coatings and single wall coating. This is because the removal efficiency is limited by the reaction rate rather than the diffusion rate, as observed and analyzed previously. In contrast, for a high N-TiO$_2$ dosage, the removal efficiency is slightly improved by increasing the room temperature for both coating layouts. This is because that diffusion is the limiting factor under this condition. Moreover, the toluene removal efficiency is improved more for the single wall coating than that for the separate surface coatings. Increasing temperature from 5 °C to 35 °C, the time required to remove 80 % of the toluene is reduced to three days for the single wall coating, while it is reduced to less than one and a half day for the separate surface coatings. This is due to a longer diffusion distance for the single wall coating. As a result, increasing diffusion velocity plays a more important role in improving the VOC removal efficiency.

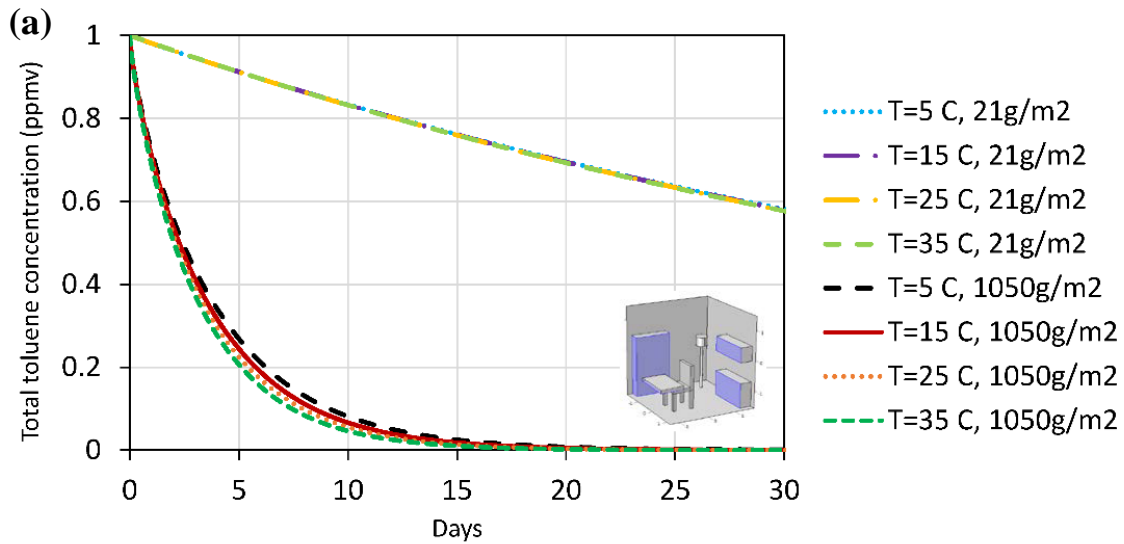

(a)

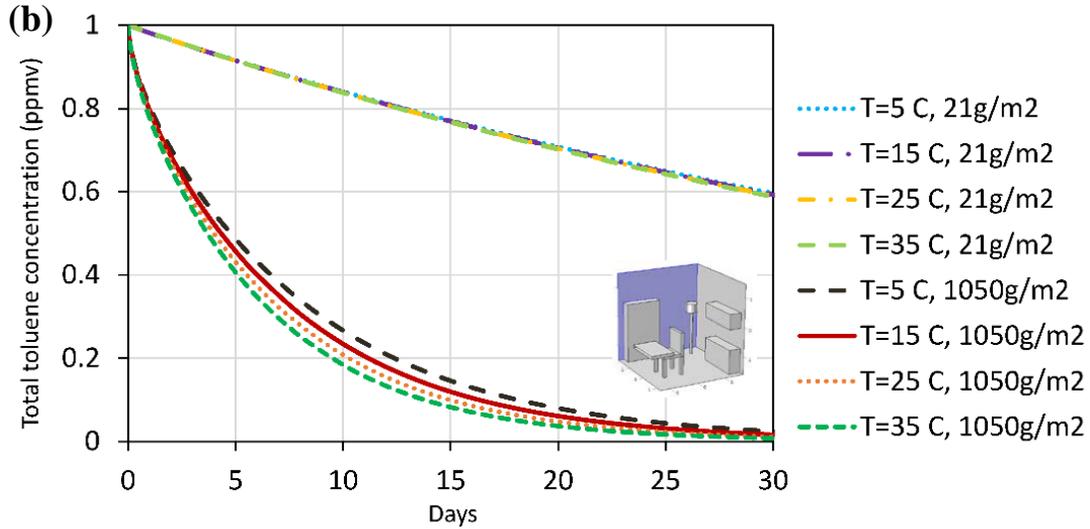

Figure 5: Toluene removal efficiency under different room temperatures.

(a) Separate surface coatings. (b) Single wall coating.

### 3.3 Accelerating contaminant migration with a fan

The second strategy to improve contaminant removal efficiency studied is to accelerate contaminant migration onto the photocatalytic coatings by using a fan. The operation of a fan can accelerate the air circulation inside the room and feed air contaminant onto the photocatalytic coatings. Fig. 6 shows the velocity field that generated by the fan.

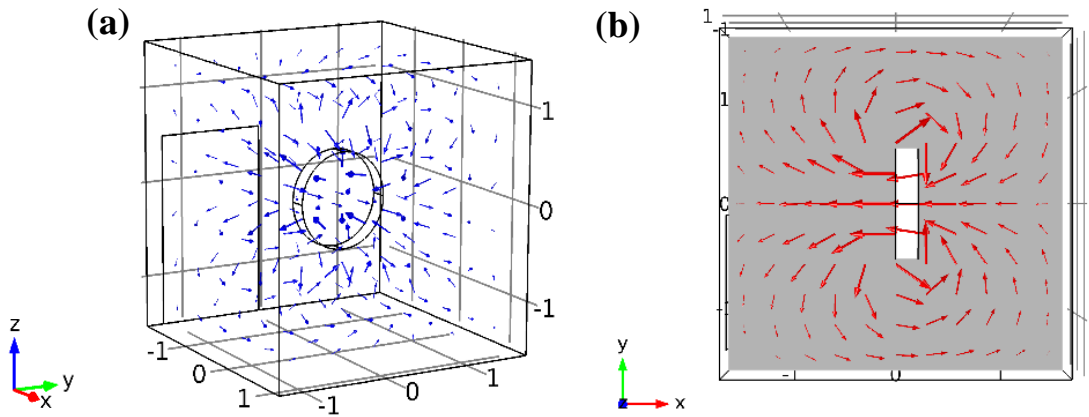

Figure 6: Velocity field generated by fan, v=0.01m s$^{-1}$. (a) 3D view, and (b) 2D view from x-y plane, z=0.

The simulation results indicate that as the fan speed increases, the removal efficiency improves, and finally the removal efficiency reaches a limit as shown in Fig. 7. At this limit, the fan makes the toluene concentration relatively uniform at each time step. Reaching this critical value of the fan speed of 0.01 m s$^{-1}$, the toluene concentration is rather uniform at any time, and the removal efficiency is not able to be further improved with the further increases in the fan speed. The speed of air circulation fan in practice is beyond this limit and it is in the range of several meters per second [38]. Fig. 8 reveals the removal efficiencies at different N-TiO$_2$ dosages with and without the use of a fan. The fan is assumed to operate above the critical speed for different N-TiO$_2$ dosages. It was found that the removal efficiencies can be all significantly improved by using a fan. The higher dosage of N-TiO$_2$, the more significant improvement of removal efficiency is achieved. With a higher dosage, there is a larger concentration difference of toluene within the room resulting from the relatively slow diffusion speed compared with photocatalytic reaction rates. Forced convection by the fan can quickly circulate the air and make toluene uniformly distributed. This process helps to feed toluene onto the photocatalytic surface, and greatly improve the toluene removal efficiency. For the highest N-TiO$_2$ dosage, the time required to remove 80 % of the toluene reduces from 10 days to 2 days by using a fan for air circulation.

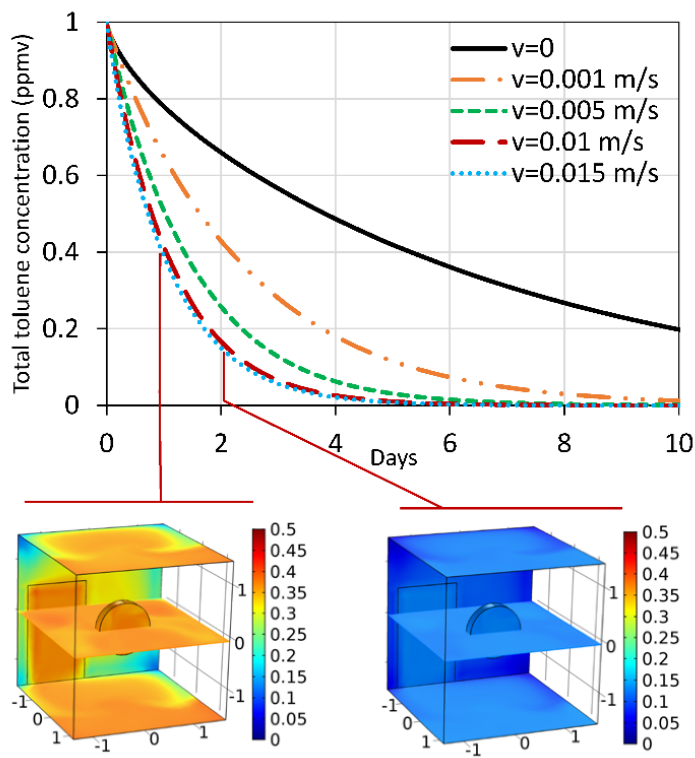

Figure 7: Toluene removal efficiency with different fan speeds, N-TiO$_2$ dosage is 1050 g m$^{-2}$.

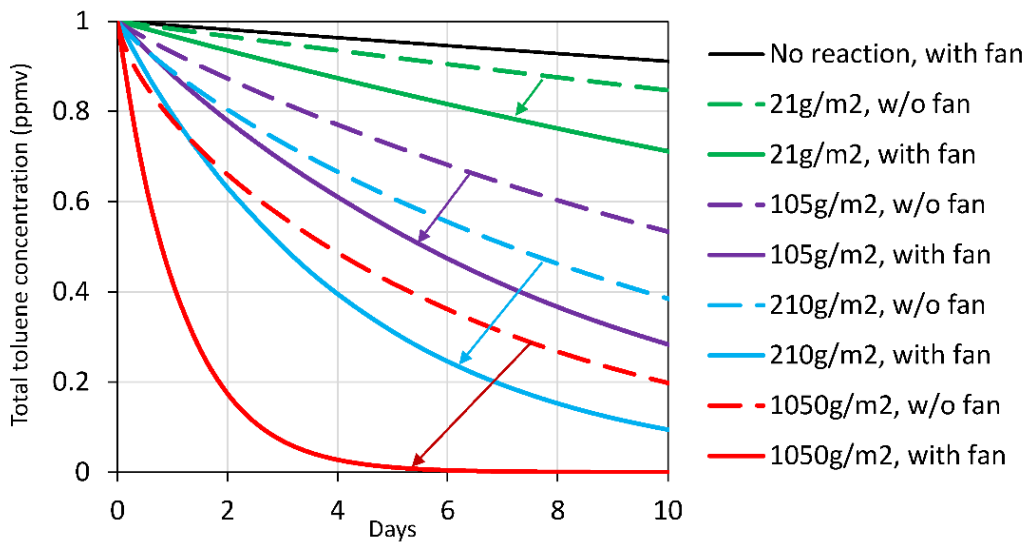

Figure 8: Comparisons of the toluene removal efficiency with and without a fan under different N-TiO$_2$ dosages.

## 3.4 Long-term functionality considering contaminant releasing

Finally, the long-term performance is analyzed by considering VOC releasing from the furniture. Two releasing rates were calculated. The passive releasing source, e.g. carpet, can slowly and continuously emit VOCs to the indoor air. The active releasing source, e.g. printer, releases VOCs abundantly when it operates. The toluene releasing rates of the two sources are listed in Table 5.

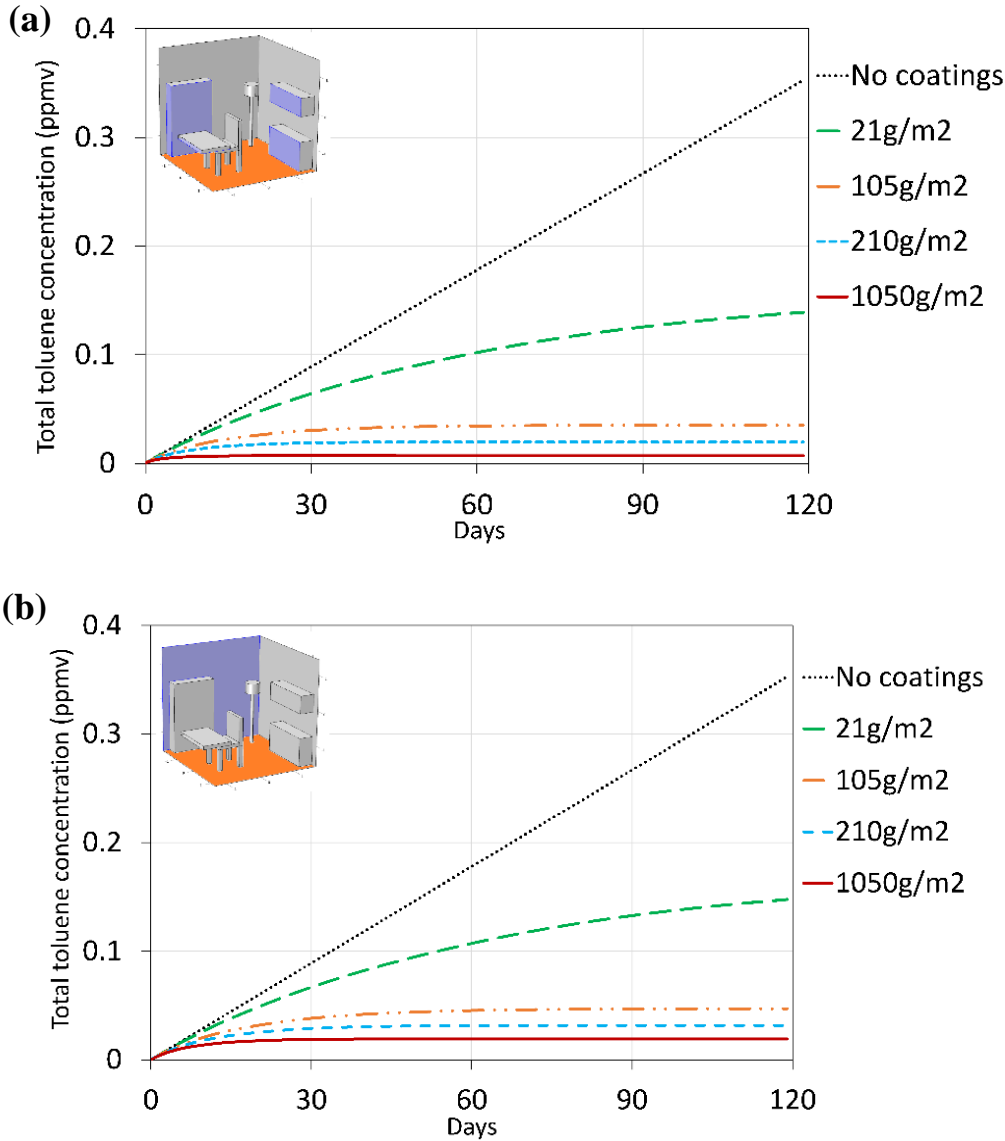

Figure 9: Toluene accumulation with a low toluene releasing rate for different N-TiO$_2$ dosages. (a) Separate surface coatings, and (b) Single wall coating.

The toluene accumulation within the room with releasing from carpet for four months is shown in Fig. 9. The black dot line represents the emitted toluene accumulation without any treatment. By using photocatalytic coatings, the toluene concentration increases, and then gradually reaches an equilibrium concentration. This equilibrium results from the balance among toluene releasing, diffusion and photodegradation on the photocatalytic surfaces. Comparing Fig. 8 (a) and (b), with the same releasing rates and N-TiO$_2$ dosages, the total toluene concentration is lower with photocatalyst applied to separate surface coatings than applied to a single wall coating. This is due to shorter diffusion distance associated with the separate surface coatings. The toluene concentrations are effectively maintained under the safety limit of 0.1 ppmv according to VOC criteria by using the N-TiO$_2$ dosage over 100 g m$^{-2}$ for both types of photocatalyst coating layouts.

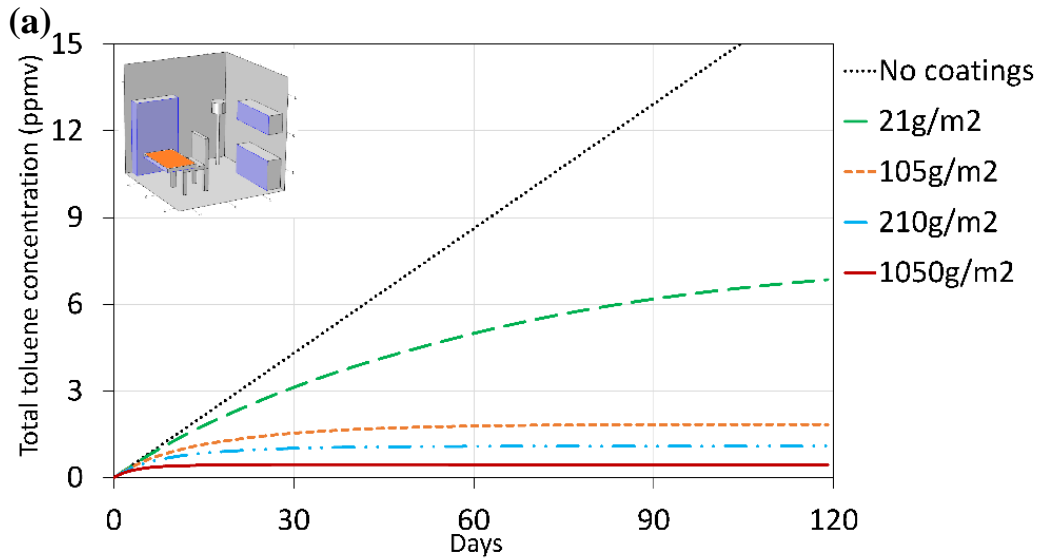

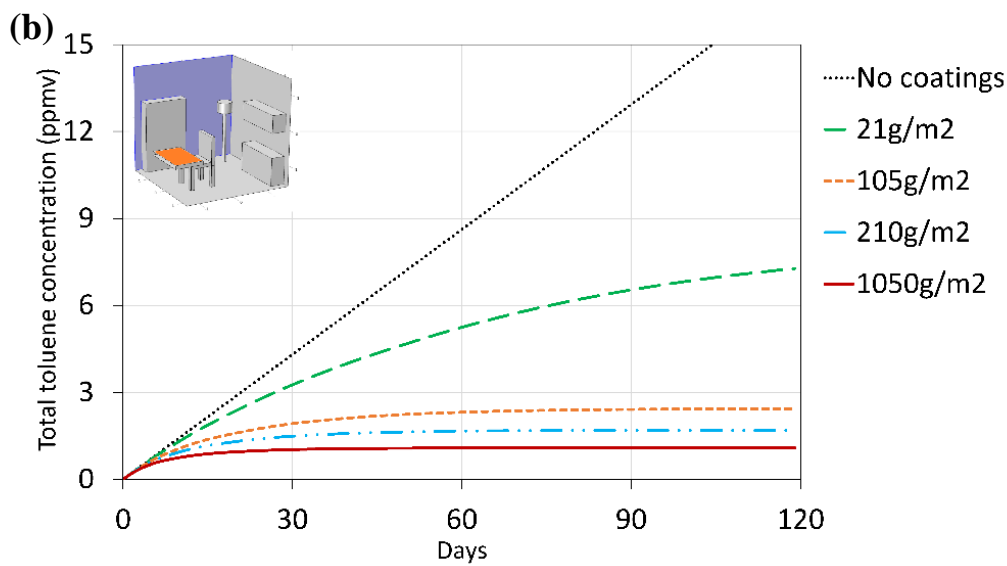

Figure 10: Toluene accumulation with a high toluene releasing rate for different N-TiO$_2$ dosages. (a) Separate surface coatings, and (b) Single wall coating.

However, for a high releasing rate (as shown in Fig. 10), the photocatalytic coatings can only maintain the toluene concentration around 1 or 2 ppmv, which does not meet the VOC criteria, even by using N-TiO$_2$ dosage of over 100 g m$^{-2}$ for both coating layouts.

### 3.5 Suggested future work prior to applications

To further assess the potential of photocatalytic coatings for practical use indoor, two questions need to be addressed. Firstly, there is a collection of VOCs in the real indoor atmosphere, and they are decomposed at different decomposition rates. Moreover, some research reported that the decomposition rate of a certain VOC is different with or without the presence of other VOCs. For example, the decomposition rate of toluene by N-TiO$_2$ was depressed by coexisting formaldehyde, while that of formaldehyde was barely influenced by coexisting toluene [39]. Among the indoor air, there are dozens kinds of VOCs, and those concentrations are varied in different buildings. They probably affect each other during photodegradation process. Therefore, the field test with a collection of VOCs are suggested to validate prior to application of the proposed photocatalyst concepts.

Secondly, intermediates can be produced during photodegradation process. Attentions should be paid that whether the intermediates are released to the surrounding air, and if yes, whether they are toxic to human being. For example, research proved that before completely decomposed to $CO_2$ and $H_2O$, toluene became benzaldehyde, carboxylic and aldehyde as the intermediates. However, these intermediates were strongly adsorbed on the photocatalyst surfaces, and no gaseous intermediates were detected in the surrounding air [40]. Therefore, it is prudent that the intermediates of other VOCs should be analyzed as well prior to practical applications.

## Conclusions

A holistic multiphysics model is developed and applied to evaluate the performance of doped-$TiO_2$ coatings activated by visible light in removing VOC contaminants. The kinetic simulations show that indoor air contaminants can be decomposed with nitrogen-doped $TiO_2$ coatings by 80 % in a room with a stagnant air condition in a week. It was also found that applying photocatalytic coatings to different areas can remove contaminants more efficiently than applying to a single location with the same total coverage area. Two strategies were proposed to improve the VOC removal efficiency. The first strategy is to increase photodegradation rate by use of a higher photocatalytic particle dosage can greatly increase the VOC removal efficiency. The second strategy is to accelerate VOC migration speed. The analyses show that increasing temperature can slightly improve VOC removal efficiency. Forced convection by a fan can facilitate indoor air circulation. It reduces the time required for 80 % contaminant removal from 10 days to 2 days. The visible light driven photocatalyst coating has a long term benefit in controlling the level of indoor air contaminants released by indoor utilities. This work demonstrates the promise of photocatalyst driven by visible light to effectively remove VOCs and create a healthy indoor environment.

**Conflict of interest**

We declare no conflict of interest.

**Acknowledgement**

This work is supported by US National Science Foundation with an award number of 1563238.